# H I and CO in the circumstellar environment of the S-type star RS Cnc*

Y. Libert[1,2], J.M. Winters[2], T. Le Bertre[1], E. Gérard[3], and L.D. Matthews[4]

[1] LERMA, UMR 8112, Observatoire de Paris, 61 Av. de l'Observatoire, 75014 Paris, France
  e-mail: `libert@iram.fr`
[2] IRAM, 300 rue de la Piscine, 38406 St. Martin d'Hères, France
[3] GEPI, UMR 8111, Observatoire de Paris, 5 Place J. Janssen, 92195 Meudon Cedex, France
[4] MIT Haystack Observatory Off Route 40 Westford, Massachusetts, USA

**ABSTRACT**

*Context.* The history of mass-loss during the AGB phase is key to understanding the stellar evolution and the gas and dust replenishment of the interstellar medium. The mass loss phenomenon presents fluctuations with a large variety of timescales and spatial scales and requires combining data from multiple tracers.

*Aims.* We study the respective contributions of the central source and of the external medium to the complex geometry of circumstellar ejecta.

*Methods.* This paper presents interferometric and single-dish CO rotational line observations along with H I data obtained for the oxygen-rich semi-regular variable RS Cnc, in order to probe its circumstellar environment at different scales.

*Results.* With the Plateau de Bure Interferometer and the IRAM 30-m telescope, we detect both the CO(1-0) and the CO(2-1) rotational lines from RS Cnc. The line profiles are composite, comprising two components of half-width $\sim 2$ km s$^{-1}$ and $\sim 8$ km s$^{-1}$ respectively. Whereas the narrow velocity component seems to originate from an equatorial disk in the central part of the CO envelope, the broad component reveals a bipolar structure, with a north-south velocity gradient. In addition, we obtain new H I data on the source and around it in a field of almost 1 square degree. The H I line is centered at $v_{LSR} = 7$ km s$^{-1}$ in agreement with CO observations. A new reduction process reveals a complex extended structure in the northwest direction, of estimated size $\sim 18'$, with a position angle ($\sim 310°$) opposite to the direction of the stellar proper motion ($\sim 140°$). We derive an H I mass of $\sim 3\,10^{-2}$ M$_\odot$ for this structure. Based on a non-spherical simulation, we find that this structure is consistent with arising from the interaction of the star undergoing mass loss at an average rate of $\sim 10^{-7}$ M$_\odot$ yr$^{-1}$ over $\sim 2\text{-}3\,10^5$ years with the interstellar medium.

*Conclusions.* We explore two related but well separated regions of the circumstellar environment around RS Cnc using CO and H I lines. With CO, we probe the recent history of mass loss that shows a bipolar geometry which is probably related to the intrinsic behavior of the mass loss process. In H I, we find a trail of gas, in a direction opposite to the proper motion of RS Cnc lending support to the hypothesis of an interaction with the interstellar medium. This work illustrates the powerful complementarity of CO and H I observations with regard to a more complete description of circumstellar environments around AGB stars.

**Key words.** stars: individual: RS Cnc – stars: mass-loss – stars: AGB and post-AGB – stars: winds, outflows – radio lines: stars – circumstellar matter

## 1. Introduction

During their evolution on the asymptotic giant branch (AGB), low- and intermediate-mass stars experience mass loss at time-dependent rates covering a large range of values ($10^{-8}$ - $10^{-4}$ $M_\odot$ yr$^{-1}$). Most of this process is expected to occur at the end of the AGB phase, and there is a large body of observations (obtained mainly in the infrared and radio range, Olofsson 2004) supporting this view. There is also evidence of mass loss at a low rate from less luminous stars, especially those in the early phase of the AGB ($\sim 10^{-9}$ $M_\odot$ yr$^{-1}$; Omont et al. 1999). Up to now it has been difficult to evaluate the relative contribution of mass loss occurring at a low rate for a long period of time, and mass loss occurring at a large rate for a short period. It requires probing circumstellar shells over large sizes in order to obtain information on the history of mass loss.

IRAS has discovered extended emission at 60 and 100 $\mu$m around several AGB stars in the solar neighborhood (Young et al. 1993a). This emission reveals dusty envelopes of large size ($\sim$ 1 pc and more), sometimes detached from the central star, that probably result from the interaction of expanding shells with the surrounding interstellar medium (ISM, Young et al. 1993b). Evolutionary models of the interaction of AGB outflows with the ISM predict large regions, up to 2.5 pc, of neutral atomic gas surrounding AGB stars (Villaver et al. 2002). Observations of the atomic hydrogen line at 21 cm with the Nançay Radio Telescope (NRT) have revealed such large circumstellar regions (Gérard & Le Bertre 2006, hereafter GL2006). These observations are sometimes difficult to interpret due to the competing emission of hydrogen in the ISM, but they can provide unique information on the kinematics and physical conditions of the gas within the external regions of circumstellar shells. For instance the H i observations of Y CVn show that its detached shell results from the slowing-down of a long-lived stellar wind by surrounding matter and that the gas is at a typical temperature of $\sim$ 200 K (Libert et al. 2007, hereafter Paper I).

Evolved stars on the AGB are moving through their local ISM sometimes at relatively large velocities (e.g. Mira at $\sim$ 130 km s$^{-1}$). Villaver et al. (2003) have performed numerical simulations of the evolution of the circumstellar environment of a low-mass star moving supersonically through its surrounding ISM. They found that the circumstellar shell is progressively distorted, and predicted the formation of a cometary structure behind the star where most of the mass ejected during the AGB phase could be stored. In their survey of H i from red giants, GL2006 found emission shifted in position and in velocity with respect to the central stars, and suggested that it could be an effect of the stellar motion relative to the ISM. Using the Very Large Array (VLA), Matthews & Reid (2007, hereafter MR2007) have imaged the H i emission from RS Cnc discovered by Gérard & Le Bertre (2003, hereafter GL2003). Their image reveals the head-tail morphology expected from the motion of the star through the ISM. The same kind of structure was found for Mira by Matthews et al. (2008), and these authors proposed that extended gaseous tails may be ubiquitous in mass losing evolved stars. The Mira's case is of special interest because, using GALEX, Martin et al. (2007) discovered a tail visible in the far-ultraviolet and extending over 2 degrees on the sky.

---

⋆ Based on observations carried out with the IRAM Plateau de Bure Interferometer. IRAM is supported by INSU/CNRS (France), MPG (Germany) and IGN (Spain).

The H i spectra obtained with the NRT, at different positions along the tail up to 2 degrees reveal a deceleration of the circumstellar gas by the local ISM. Recently, Libert et al. (2008, hereafter Paper II) found evidence of an extended gaseous tail associated with RX Lep, with physical conditions similar to those in the Y CVn detached shell.

In the present paper we revisit the case of RS Cnc in order to better document the interaction between circumstellar shells and the ISM. In addition to having an already known cometary morphology in H i, this source appeared particularly well suited for such study because, from observations in the CO rotational lines, there is evidence of a long term variation in the characteristics, such as the expansion velocity and the mass-loss rate, of its outflow (Knapp et al. 1998, hereafter K1998) as well as of a bipolarity in the inner molecular shell (Neri et al. 1998). By combining CO and H i results, we can thus explore the respective effects of possibly interacting wind, and of asymmetric mass loss, on the properties of the outer circumstellar shell. The complementarity of CO and H i data was already profitably exploited in our study of the circumstellar environment of RX Lep (Paper II).

## 2. Observational results

### 2.1. RS Cnc

RS Cnc is an oxygen-rich late-type giant (HR 3639; M6IIIase) that shows an excess of heavy s-process elements (Smith & Lambert, 1986) and Tc lines in its optical spectrum (Lebzelter & Hron 1999). It is an intrinsic S star (CSS 589) in the thermally pulsing phase of the AGB. Using the FRANEC stellar evolution code, and fitting the abundances determined by Smith & Lambert, Busso & Palmerini (2009) estimate that it is presently a $1.2\,M_\odot$ star (with $M_{ZAMS} \sim 1.5\,M_\odot$), in its $\sim$ 20th thermal pulse (10th with dredge-up). It is a semi-regular variable of type SRc with periods of order 130 and 250 days (Adelman & Dennis 2005; Howarth 2005). The effective temperature is $\sim 3200\,K$ (Dumm & Schild 1998; Dyck et al. 1996; Perrin et al. 1998). With a temperature much higher than 2500 K, and following Glassgold & Huggins (1983), we expect that most of its hydrogen should be in atomic form from the stellar atmosphere outwards.

The parallax measured by Hipparcos is $8.21 \pm 0.98$ mas (Perryman et al. 1997) which translates into a distance of 122 pc that has been adopted in different recent works. The New Reduction of the Hipparcos Data yields $6.97 \pm 0.52$ mas (van Leeuwen 2007). Hereafter we will keep the former estimate, with the caveat that the distance could in fact be slightly larger (by 10-20%). The proper motion, corrected for the solar motion towards the apex, is 20 mas yr$^{-1}$ in Right Ascension (RA) and -21 mas yr$^{-1}$ in declination (DEC). RS Cnc is therefore moving, in the plane of the sky, southeast (PA = 137°) at 17 km s$^{-1}$. With a radial velocity, $V_{LSR} = 7.5$ km s$^{-1}$ (see Sect. 2.2), we estimate the 3-D velocity of RS Cnc at 18.6 km s$^{-1}$.

RS Cnc was found to be extended in the IRAS data at 60 $\mu$m with evidence for a detached shell ($R_{in}$ = 1.0′ or 0.036 pc at 122 pc, and $R_{out}$ = 5.8′ or 0.21 pc; Young et al. 1993a). Furthermore evidence for present mass loss is given by the detection of a silicate feature in emission at 10 $\mu$m (IRAS, Speck et al. 2000).

RS Cnc was a target in many radio-line surveys. Up to now only rotational lines of CO (Nyman et al. 1992) and the 21-cm H i line (GL2003) have been detected. To date, there has been no reported detection of radio continuum emission (however, see below Sect. 2.3).

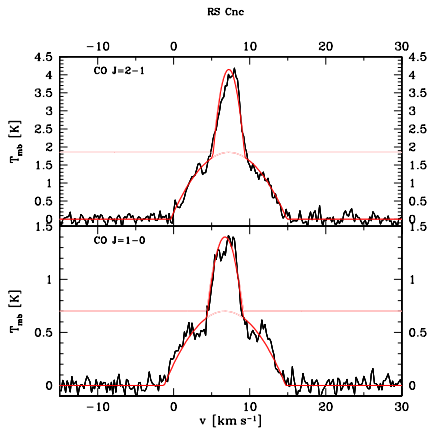

**Fig. 1.** 30-m spectra centered on RS Cnc. The fit of a two-wind model is shown in red (see Sect. 2.2). The abscissas are in LSR velocity and the spectral resolution is smoothed to $0.2\,\mathrm{km\,s^{-1}}$. The $T_{mb}$ - Jy conversion factor is $5.3\,\mathrm{Jy\,K^{-1}}$ (at 230 GHz) and $4.8\,\mathrm{Jy\,K^{-1}}$ (at 115 GHz)

## 2.2. Rotational lines of CO

The CO(1-0) emission from RS Cnc was first detected by Nyman et al. (1992) with a line centered at $V_{LSR} = 7.5\,\mathrm{km\,s^{-1}}$. The source was then observed by K1998, using the 10.4 m telescope of the Caltech Submillimeter Observatory, thus providing high spectral resolution profiles for both the CO(2-1) and the CO(3-2) lines. In these spectra, the lines are also centered at $7.5\,\mathrm{km\,s^{-1}}$ and exhibit a composite profile made of a narrow feature, of half-width $\sim 2.6\,\mathrm{km\,s^{-1}}$, superimposed on a broader one of half-width $\sim 8.0\,\mathrm{km\,s^{-1}}$. K1998 interpret these profiles as produced by two successive winds with different expansion velocities (respectively $2.6\,\mathrm{km\,s^{-1}}$ and $8\,\mathrm{km\,s^{-1}}$) and different mass-loss rates ($2.3\,10^{-8}\,M_\odot\,\mathrm{yr}^{-1}$ and $1.0\,10^{-7}\,M_\odot\,\mathrm{yr}^{-1}$). This type of profile is not unique among AGB stars, though it seems more frequent among semiregular variables than among Miras. An explanation of this phenomenon, as proposed in K1998, is that a change in the wind properties could occur when the central star undergoes a major change in its properties (e.g. luminosity, pulsation mode or chemistry), the slow wind being the mark of the onset of a new mass loss phase.

Neri et al. (1998) also observed RS Cnc in CO(1-0) and (2-1) using the IRAM Plateau de Bure Interferometer (PdBI) combined with the IRAM 30-m telescope. The observations were performed in October 1990 using the three Bure antennas available at that time. The authors find an extended CO shell, with a size of about $10''$. The position-velocity diagrams indicate that the envelope is not spherically symmetric and suggest a bipolar geometry. Nevertheless, their spatial resolution ($\sim 7''$ in 1-0), as well as their spectral resolution ($1.3\,\mathrm{km\,s^{-1}}$ in 2-1 and $2.6\,\mathrm{km\,s^{-1}}$ in 1-0), are not sufficient to characterize this phenomenon in more detail.

Thus, we have re-investigated the CO(1-0) and CO(2-1) emission of RS Cnc with the PdBI and the 30-m telescope with the same observational strategy as for EP Aqr (Winters et al. 2007), a semiregular variable of M-type which also shows composite CO line profiles (K1998, Winters et al. 2003). The interferometric data were obtained between November 2004 and April 2005 using 6 antennas in 3 configurations (B, C and D; with a range of baselines from 24 m in D-configuration up to 330 m in B-configuration) for a total of 20 hours of integration time. To recover the short spacing information and restore the extended emission filtered out by the interferometer, we obtained on-the-fly (OTF) maps centered on the stellar position of RS Cnc at the IRAM 30-m telescope. A region of 100″ × 100″ was covered. Each OTF map consisted of 21 rows separated by 5″; each scan includes 25 dumps giving a spatial increment in the grid of 4″. The area has been covered 13 times to achieve a signal-to-noise ratio of ∼ 5 for the broad component. For both PdBI and 30 m radio telescope, the data were obtained at a spectral resolution of 0.1 km s$^{-1}$.

**Table 1.** Results of the 30-m CO spectral line fitting. Formal uncertainties are given in parentheses.

|  |  | $v_c$ (km s$^{-1}$) | $v_{exp}$ (km s$^{-1}$) | $T_{mb}$ (K) | $\dot{M}$ ($10^{-7}$ M$_\odot$ yr$^{-1}$) | $r_{CO}$ ($10^{16}$ cm) |
|---|---|---|---|---|---|---|
| (1-0) | broad | 6.75 (0.25) | 8.00 (0.5) | 0.70 (0.05) | 4.4 (0.6) | 3.5 (0.2) |
|  | narrow | 6.75 (0.25) | 2.4 (0.1) | 0.70 (0.05) | 0.86 (0.08) | 2.2 (0.1) |
| (2-1) | broad | 7.25 (0.25) | 7.65 (0.5) | 1.85 (0.1) | 3.4 (0.4) | 3.1 (0.2) |
|  | narrow | 7.25 (0.25) | 2.1 (0.1) | 2.3 (0.1) | 0.64 (0.06) | 2.0 (0.1) |

The results of the 30-m observations (Fig. 1) already give some important parameters of the wind. The averages of the spectra around the center position in the map present a two component profile as observed by K1998 and Neri et al. (1998). We find an expansion velocity of 8 km s$^{-1}$ for the CO(1-0) broad component, and 2.4 km s$^{-1}$ for the narrow one. Both features are centered on 6.75 km s$^{-1}$. For the CO(2-1) line profile, we mesure expansion velocities of 7.65 km s$^{-1}$ for the broad component, and 2.1 km s$^{-1}$ for the narrow component, respectively. The features are centered on 7.25 km s$^{-1}$. We apply the method described in Winters et al. (2003, their Sect. 4.3), using the two expressions given by Loup et al. (1993), in order to estimate the mass-loss rate and the CO photo-dissociation radius, $r_{CO}$, using the line fitting shown in Fig. 1. This method relies on the modelling of CO lines by Knapp & Morris (1985) in the case of optically thick emission. However, as already noted by Knapp & Morris, the CO(1-0) line emission in RS Cnc is probably optically thin. Winters et al. (2003) found a systematic difference by a factor ∼ 3.5 between their estimates and those of Olofsson et al. (2002) and Schöier & Olofsson (2001), that are based on a detailed modelling of several CO lines, for stars with mass loss rates around 1 10$^{-7}$ M$_\odot$ yr$^{-1}$ (figure 4 in Winters et al. 2003). Our values in Table 1 may thus overestimate the mass loss rate by a factor ∼ 3.5. A CO photodissociation radius of ∼ 3 10$^{16}$ cm would also be consistent with a low mass loss rate of ∼ 1 10$^{-7}$M$_\odot$ yr$^{-1}$ (Mamon et al. 1988, their figure 3). Finally, we note that Winters et al. (2003) derived mass loss rates from CO(2-1) line profiles by scaling them to the CO(1-0) ones. Therefore our CO(2-1) mass loss rate estimates for RS Cnc are redundant with those directly obtained from CO(1-0). On the other hand, the agreement in the results does not bring evidence of mass loss variation in RS Cnc such as those discovered by Kemper et al. (2003) and Teyssier et al. (2006), who needed to invoke such changes in the mass loss rate on timescales of a few hundred years in order to reconcile CO data obtained at low and high rotational transitions.

The images produced after merging the PdBI and the 30-m data and subtracting the continuum are shown in Fig. 2. For the CO(1-0) emission, the field of view is 44″ and the beam has a size of 2.61″ × 1.98″, oriented at a PA of 51°. For the (2-1) transition, the field of view is 22″ and the size of the beam is 1.38″ × 0.97″, oriented at a PA of 27°. Nevertheless, we adopted a common field of view of 35″ × 35″ to allow better comparison between both transitions.

In Fig. 3, we present azimuthally-averaged intensity profiles based on the CO channel maps centered at $V_{LSR}$ = 7 km s$^{-1}$. The CO(1-0) shell extends to ∼ 12″ (2.2 10$^{16}$ cm) in radius, whereas the CO(2-1) shell extends to only ∼ 9″ (1.7 10$^{16}$ cm). The difference is likely due to differences in excitation conditions for the two transitions such that the higher transition comes from a smaller region. The above shell extensions are defined at the $3\sigma$ noise level in the corresponding maps. We integrate the fluxes within the same area as Neri et al. (1998) and we obtain 117 Jy × km s$^{-1}$, and a peak at 15 Jy for the integrated CO(1-0) spectrum, and 700 Jy × km s$^{-1}$, and a peak at 100 Jy for the CO(2-1). We derive slightly higher fluxes than Neri et al. (1998) who got peaks of 9.5 Jy in (1-0) and 89 Jy in (2-1).

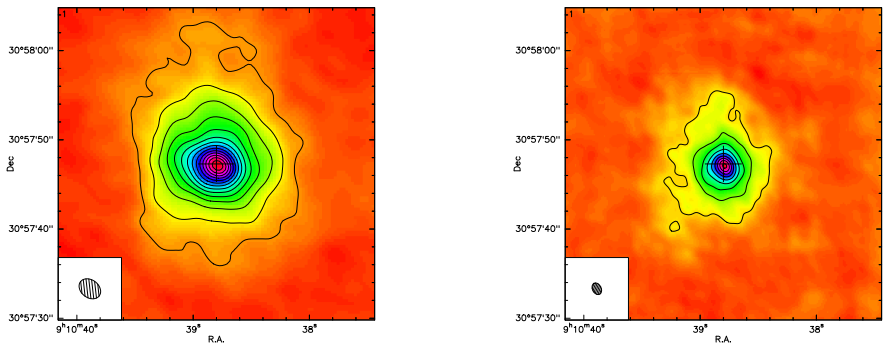

**Fig. 2.** CO(1-0) (left panel) and CO(2-1) (right panel) emission of RS Cnc integrated over the width of the line (from -2 km s$^{-1}$ to 17 km s$^{-1}$). The beam size and its position angle are $2.61 \times 1.98''$ and $51°$ for the CO(1-0) line, and $1.38 \times 0.97''$ and $27°$ for the CO(2-1) line. The contour levels range from 0.5 to 7.5 Jy beam$^{-1}$ × km s$^{-1}$ with a step of 0.5 Jy beam$^{-1}$ × km s$^{-1}$ for the CO(1-0), and from 2 to 18 Jy beam$^{-1}$ × km s$^{-1}$ with a step of 2 Jy beam$^{-1}$ × km s$^{-1}$ for the CO(2-1). Those maps were produced by merging PdBI and 30-m data.

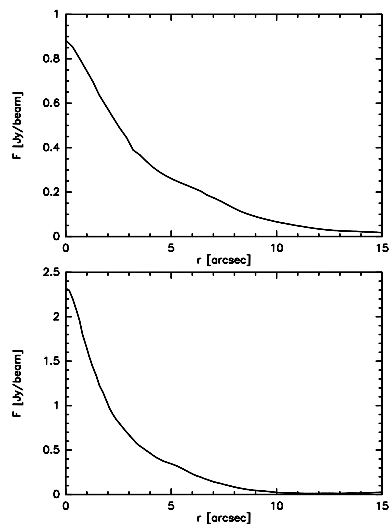

**Fig. 3.** Azimuthal average of the CO(1-0) (upper panel) and CO(2-1) (lower panel) brightness distribution in the central channel ($v_{LSR} = 7$ km s$^{-1}$), of width of 0.8 km s$^{-1}$.

The channel maps for the CO(1-0) and (2-1) lines are presented in Figs. 4 and 5 . To increase the quality of the spectra, we smoothed them to a resolution of 0.4 km s$^{-1}$. In the (1-0) line, blue-shifted emission is seen extended and first north of the stellar position (from $-0.6$ to 0.6 km s$^{-1}$), then is seen still north but peaking on the star (0.6 to 4.2 km s$^{-1}$). It then switches south, coming back on the stellar position at 6.6 km s$^{-1}$. At this velocity the image is seen elongated at a PA of about $100°$ (see also in Fig. 6 the lower left panel). From 7.0 to 8.6 km s$^{-1}$, the emission shows an extension north. Then, it moves south, peaking on the star from 9.0 to 11.8 km s$^{-1}$. And finally, from 11.8 to 14.2 km s$^{-1}$, the emission peaks south of the stellar position. The visual inspection of these channel maps reveals a symmetry with respect to 6.6 km s$^{-1}$, which is close to the central velocity determined from the line fitting (Table 1). The inspection of the (2-1) channel maps reveals the same features but less marked. The images are more symmetric, although slight offsets in position can be noted that are consistent with those observed in (1-0). Also the symmetry is rather around 7.0-7.4 km s$^{-1}$ than around 6.6 km s$^{-1}$, a trend consistent with the (2-1) line fitting.

The position-velocity diagrams (Figs. 6 & 7) complement this description. In CO (1-0), the velocity-DEC diagram (Fig. 6) shows clearly 2 opposite S-shaped features, a first one between 5 and 9 km s$^{-1}$, and a second one between 0 and 14 km s$^{-1}$. In the RA-velocity diagram, the first feature has a counterpart, also S-shaped, between 6 and 7 km s$^{-1}$. On the other hand, in this diagram, the counterpart of the second feature appears to line-up at the RA of the star. In both diagrams the intensity of this second feature peaks at 1 km s$^{-1}$ and 12.5 km s$^{-1}$. In the CO(2-1) position-velocity diagrams (Fig. 7), the same kind of features are observed, although less conspicuously.

The inspection of Figs. 4 to 7 readily reveals several features:

- The CO(1-0) and CO(2-1) maps in Figs. 4 and 5 are not consistent with a spherical envelope which would produce circular contour levels of increasing radius with decreasing velocity difference from the systemic velocity (e.g. IRC+10216: Fong et al. 2006, their Fig. 1).
- The CO(1-0) and CO(2-1) maps of Figs. 4 to 7 are different and the former are more asymmetric than the latter likely due to different excitation conditions throughout the envelope. This is a well known effect in carbon stars with detached shells (eg. Olofsson et al., 1996) where the CO(1-0)/(2-1) line ratio is larger in the outer shells than in the inner shells.
- The double S-shaped morphology of the velocity-position (DEC) CO(1-0) diagram in Fig. 6 suggests an envelope made of two main components with expansion velocities of 7 km s$^{-1}$ and 2 km s$^{-1}$, respectively.

In summary, the CO(1-0) data strongly suggest that the broad velocity component traces a bipolar outflow and the narrow velocity component, a flaring expanding disk perpendicular to the bipolar outflow and slightly tilted at PA ∼ 100°. Such an axisymmetrical structure has already been observed in proto-planetary nebulae (Bujarrabal et al. 2005) and even in some AGB stars (Chiu et al. 2006).

### 2.3. Continuum emission at 1.3 and 2.6 mm

We detect an unresolved continuum source with a flux density of 5.4 mJy at 2.6 mm (rms noise 0.3 mJy) and 16 mJy at 1.3 mm (rms noise 0.4 mJy), centered on the stellar position of RS Cnc. Adopting an effective temperature of 3226 K and a radius $R_\star$ of 192 $R_\odot$ as given by Dumm & Schild (1998), and assuming a distance of 122 pc, the black body law results in fluxes of 5.3 mJy at 2.6 mm, and 21 mJy at 1.3 mm. Thus, there is no evidence of an excess of radio emission in addition to the stellar photosphere, and an optically thick radio photosphere at 2 $R_\star$ (e.g. Reid & Menten, 1997) is excluded for this source, similar to EP Aqr (Winters et al. 2007).

### 2.4. H I observations

RS Cnc was first detected in the H I line at 21 cm by GL2003 using the NRT. After Mira (Bowers & Knapp 1988), it was the second AGB star detected in emission at 21-cm. The NRT is a merid-

8     Libert et al.: H I & CO in RS Cnc

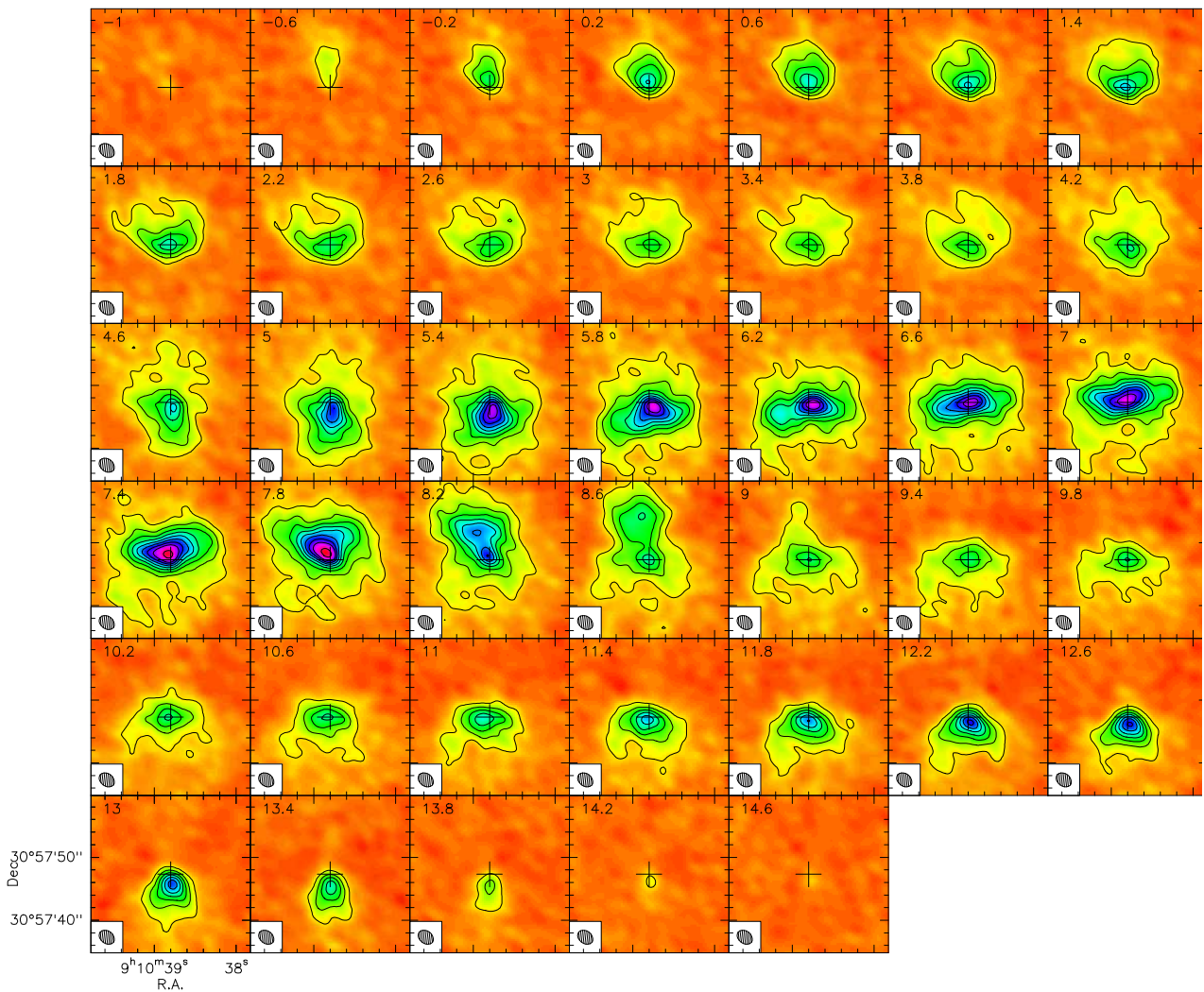

**Fig. 4.** Channel maps (PdBI+30 m) of the CO(1-0) emission. The spectral resolution is smoothed to 0.4 km s$^{-1}$. The contour levels range from 0.1 to 1 Jy beam$^{-1}$ with a step of 0.1 Jy beam$^{-1}$. The beam size and its position angle are 2.61 × 1.98″ and 51°.

ian telescope with a rectangular primary mirror of effective dimensions 160×30 m. At 21 cm, this translates to an angular resolution of 4′ in RA and 22′ in DEC. The data were obtained mostly in the position-switch mode with off-positions at ± 4′, ± 6′, ± 8′, and ± 12′ in RA. Despite a strong contamination around -8 km s$^{-1}$ LSR due to the interstellar H I emission, the source was detected close to the expected velocity. The line profile obtained by averaging all the position-switched spectra is composite and reminiscent of the CO rotational line profiles. A narrow rectangular component of width 4 km s$^{-1}$ appeared over a broad quasi-gaussian component of width (FWHM) ~ 12 km s$^{-1}$. GL2003 interpreted this detection in support of the Glassgold & Huggins (1983) model which predicts that atomic hydrogen should dominate in circumstellar environments of late-type stars with effective temperature larger than 2500 K.

Using the VLA, MR2007 found that the H I source has a compact feature centered on the star (Fig. 8), plus a filament extending ~ 6′ to the northwest (PA ~ 310°). This morphology suggests a physical association with RS Cnc and MR2007 concluded that the observed atomic hydrogen originates in the stellar atmosphere.

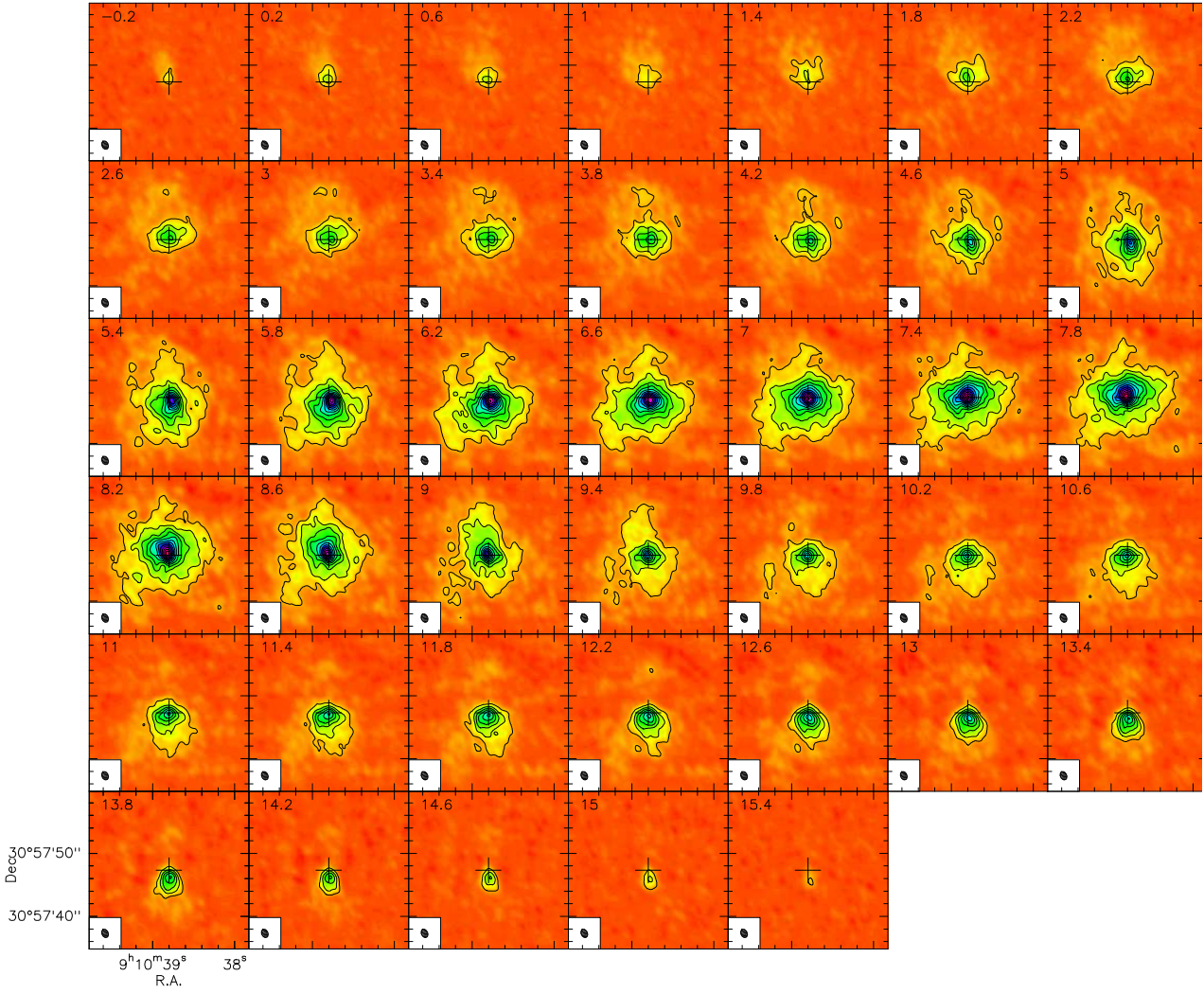

**Fig. 5.** Channel maps (PdBI+30 m) of the CO(2-1) emission. The spectral resolution is smoothed to $0.4\,\mathrm{km\,s^{-1}}$. The contour levels range from 0.2 to $2.4\,\mathrm{Jy\,beam^{-1}}$ with a step of $0.2\,\mathrm{Jy\,beam^{-1}}$. The beam size and its position angle are $1.38 \times 0.97''$ and $27°$.

From this recent work it is clear that the source detected by GL2003 is much larger than the NRT beam in RA and that its own flux contributed to the reference positions at $-4'$, $-6'$, and $-8'$, resulting in a decrease of the total measured flux and possibly in a deformation of the line profile obtained by GL2003. In view of the particular interest of RS Cnc, and of this problem, we have obtained more H i data with the NRT, covering a much larger region around RS Cnc in order to allow a better separation between the genuine stellar emission and the ambiant Galactic H i emission. We analyzed the complete body of data with a new approach taking into account the northwest extension of the source discovered by MR2007.

The whole set of the NRT data including the most recent observations of RS Cnc represents a total of 118 hours of observing time, between May 2002 and October 2008. Data were obtained using the position-switch technique with two off-positions (east and west) and a beam throw up to $32'$. The vicinity of RS Cnc was sampled every half beam in RA and in DEC. Each spectrum has a bandwidth of $165\,\mathrm{km\,s^{-1}}$ and a channel width of $0.08\,\mathrm{km\,s^{-1}}$. For convenient analysis, we smoothed the spectra with a Hanning filter down to a spectral resolution of $0.16\,\mathrm{km\,s^{-1}}$. The data were processed with the CLASS software, part of the GILDAS[1] package developed at IRAM (Pety 2005).

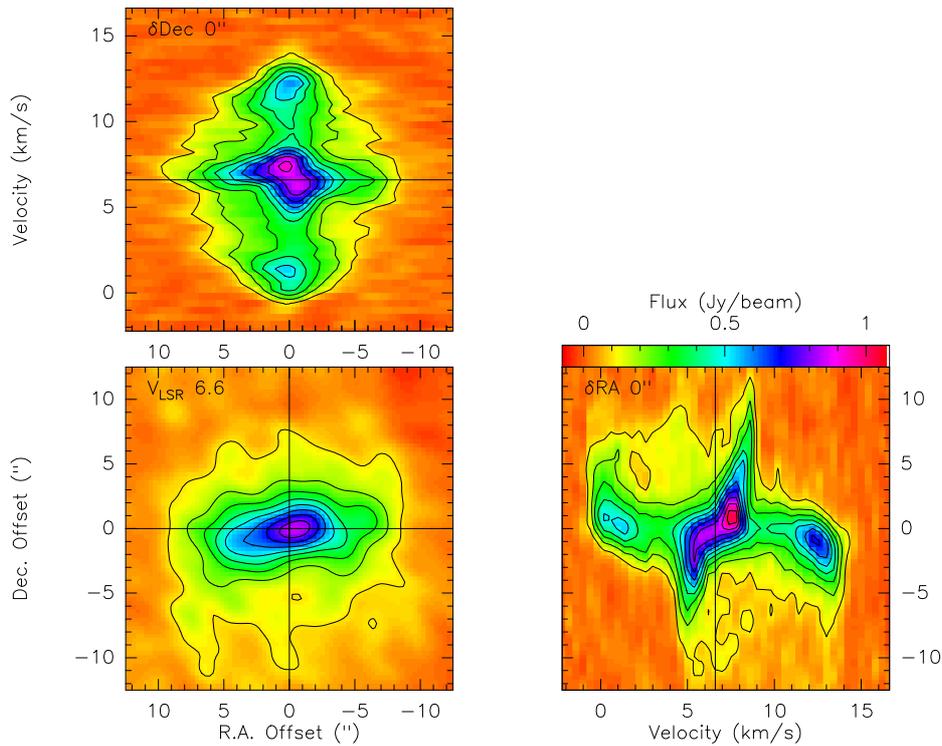

**Fig. 6.** Lower left panel: CO(1-0) channel map at 6.6 km s$^{-1}$. Upper left: position (RA)-velocity diagram at the declination of the star. Lower right: velocity-position (DEC) diagram at the RA of the star. The contour levels range from 0.1 to 0.8 Jy beam$^{-1}$ with a step of 0.1 Jy beam$^{-1}$.

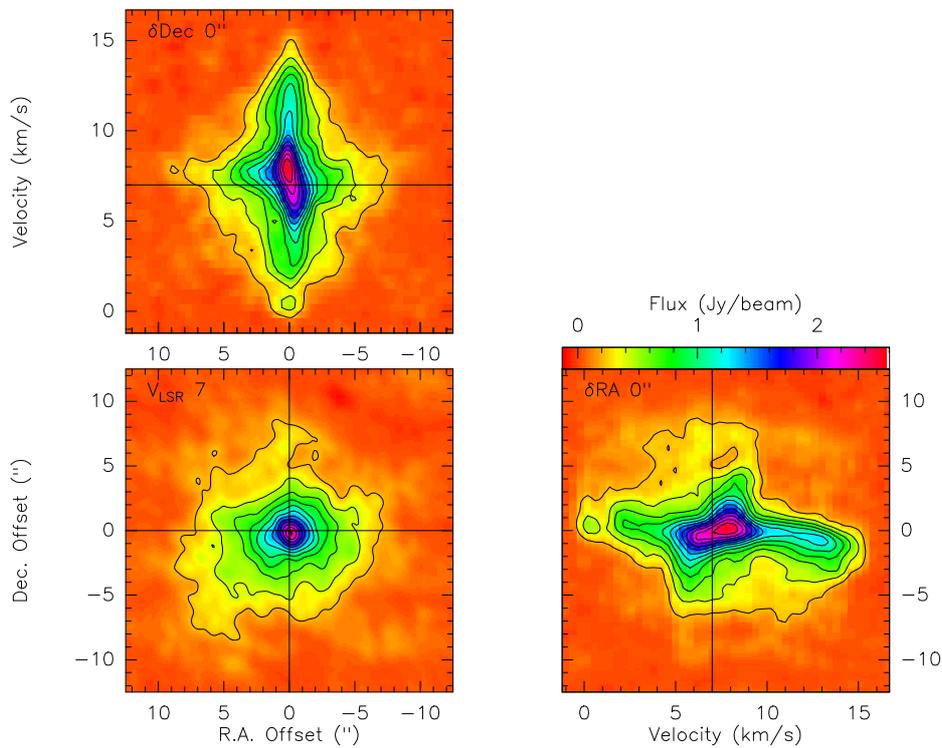

**Fig. 7.** Lower left panel: CO(2-1) channel map at 7 km s$^{-1}$. Upper left: position (RA)-velocity diagram at the DEC of the star. Lower right: velocity-position (DEC) diagram at the RA of the star. The contour levels range from 0.2 to 2.4 Jy beam$^{-1}$ with a step of 0.2 Jy beam$^{-1}$.

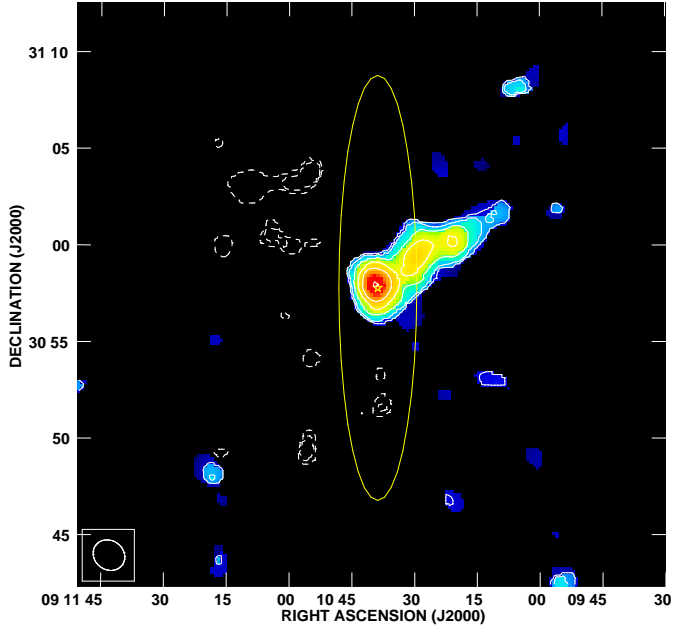

**Fig. 8.** H I emission of RS Cnc obtained with the VLA (MR2007). The NRT beam is represented in yellow. The intensity range is 0-100 Jy beam$^{-1}$ × m s$^{-1}$ and the contours are (-16, -8, 8, 16, 24, 32, 48, 64, 88) × 1.25 Jy beam$^{-1}$ × m s$^{-1}$.

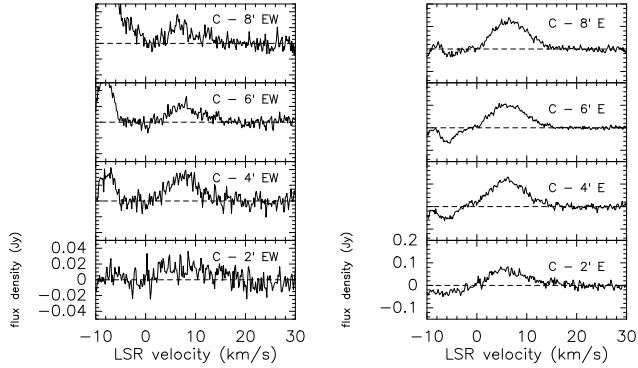

**Fig. 9.** RS Cnc position-switched spectra. The left panel shows spectra obtained using both the east and west offset position. The right panel displays the spectra obtained with the east comparison only.

Following the method described in Paper II, we revisit the analysis of the H I data of RS Cnc. Fig. 9 (left panel) shows the position-switched spectra obtained with the two east and west off-positions averaged and subtracted from the ON profile spectrum. If the offset spectra were free from source emission, the resulting lines should reach a maximum intensity as the distance of the offset position increases. But, Fig. 9 (left panel) clearly shows that, instead of converging toward a maximum intensity, the peak decreases. This means that at least one of the offset references is contaminated either by the source itself, or by the Galactic H I emission. Fig. 9 (right panel) shows the results of the position-switched spectra obtained by subtracting the east reference only. In that case, the intensity reaches a maximum around 4′ east. Consequently, we can consider that the east side of the source is not polluted by source emission beyond 4′ east, and we choose only the eastward reference to process the map. The average of the spectra, with off-positions selected only beyond 4′ to the east, is adopted as the center spectrum (cf. Fig. 10).

---
[1] http://www.iram.fr/IRAMFR/GILDAS

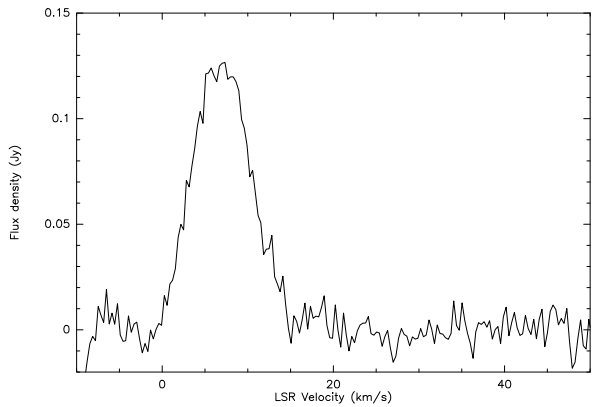

**Fig. 10.** H i emission of RS Cnc obtained with the NRT new set of data at the stellar position.

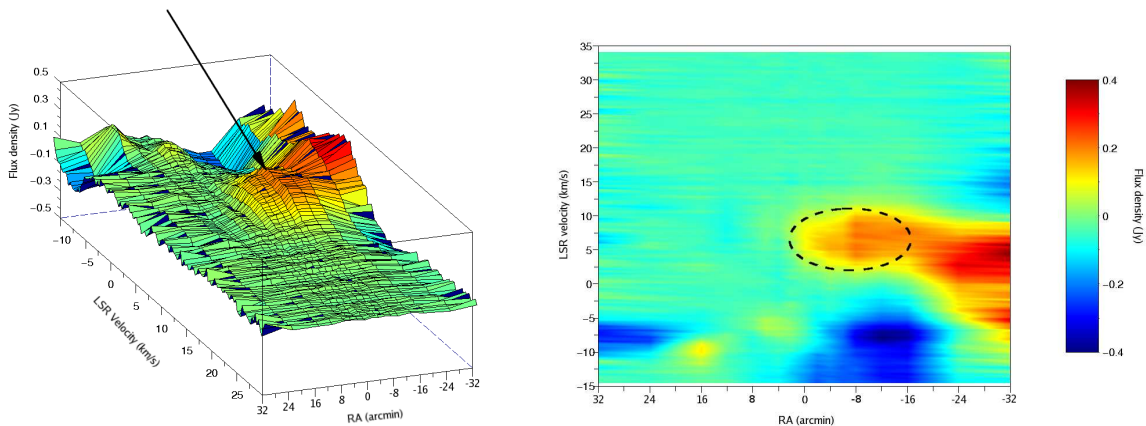

**Fig. 11.** Left panel: position (RA)-velocity-flux 3D view of RS Cnc at the declination of the star. The arrow points to the H i cloud associated with RS Cnc. Right panel: the same data projected in a 2D view, the dashed ellipse delineates the assumed H i emission from the source.

Separating the genuine emission of the source from the underlying contamination is still a critical problem. Thus, we used the method described in Paper II to produce a more practical display of the H i emission around RS Cnc. The result is shown in Figs. 11, 12 and 13. These figures present the spectra for every step in RA obtained with the NRT at different declinations around the source. These views clearly confirm that the east side of RS Cnc is free of H i confusion and show that there is additional polluting emission to the west, broader than the expected emission of RS Cnc, and not exactly centered at its LSR velocity, but rather blueshifted by $\sim 2\,\mathrm{km\,s^{-1}}$. But the remarkable feature common to all of the 3D views is a "valley" located near RA = 16′ to the west, at all radial velocities, that separates the RS Cnc H i emission from the Galactic background emission. This spatial separation is most useful because the H i confusion occurs at radial velocities close to that of RS Cnc. The expected width of the line and the radial velocity of the star derived from the CO observations (Sect. 2.2) combined with this display allow us to constrain the full extent of RS Cnc and isolate the intrinsic emission (Fig. 11, dashed ellipse). Thus, we estimate that the circumstellar emission is lying between 2′ east and 16′ west, leading to a size of the H I envelope of ∼ 18′ (∼ 0.65 pc at 122 pc).

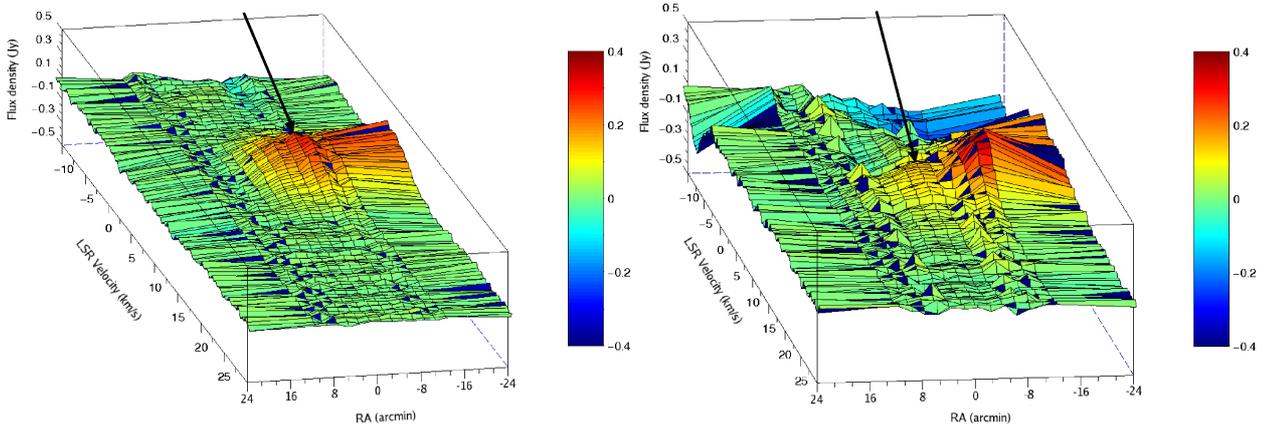

**Fig. 12.** Same as in Fig. 11, left panel, for 11′ north (left panel) and 11′ south (right panel). The arrow points to the assumed position of the RS Cnc H I emission.

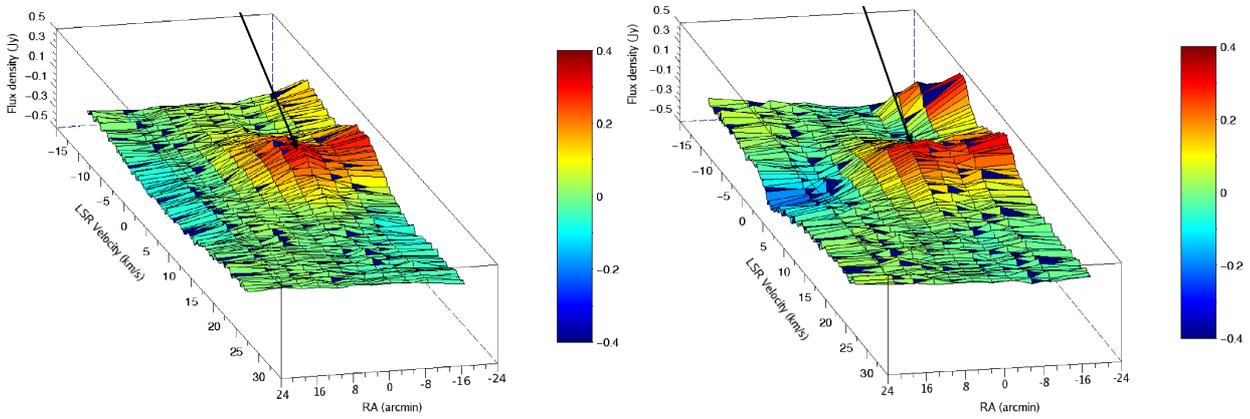

**Fig. 13.** Same as in Fig. 11, left panel, for 22′ north (left panel) and 44′ north (right panel). The arrow points to the assumed position of the RS Cnc H I emission.

Fig. 11 reveals that the H I emission of RS Cnc has been underestimated in GL2003, due to the H I emission to the west originating from the source and from the Galactic hydrogen. The center spectrum exhibits a quasi-Gaussian component, of intensity 0.129 Jy, centered at $V_{LSR} = 6.97 \pm 0.06$ km s$^{-1}$ and FWHM = $7.8 \pm 0.1$ km s$^{-1}$, in good agreement with previous CO results and our new IRAM data.

At 11′ north and 11′ south, the confusion is more critical and nearly coincides in velocity with the source, making the separation less easy. Besides, the intrinsic emission of the source appears to arise on top of some underlying Galactic emission. Thus, we need to evaluate the background emission for each set of spectra obtained at a given declination, from -11′ south to 44′ north. We fit a linear baseline to the emission beyond 2′ east and 16′ west, in the LSR velocity range of RS Cnc, i.e. over 8 km s$^{-1}$. By integrating the resulting spectra and summing them up, we derive an integrated flux of 7.7 Jy × km s$^{-1}$. Using the standard relation, $M_{HI} = 2.36\ 10^{-7} d^2 \int S_V dV$, it translates to ∼ $2.7\ 10^{-2}$ M$_\odot$ in atomic hydrogen, at 122 pc.

## 3. Interpretation

We observe CO profiles made of two components: a narrow component superimposed on a broad one, both centered at $\sim 7 \, \mathrm{km \, s^{-1}}$ (Sect. 2.2). Owing to the high spatial and spectral resolution of our new data, we are able to separate these two components. The narrow component originates mainly from an elongated structure with a PA $\sim 100°$. The broad component originates from another elongated structure, but at a PA $\sim 10°$ with the northern part moving towards us and the southern part moving away from us. Such a morphology could be explained by the combination of a bipolar outflow and an equatorial waist, or "disk".

Although this axisymmetrical structure accounts for the north/south CO extension, it does not constrain the nature of the equatorial region, which could be either a disk in expansion or in rotation. However both the upper and right panels in Fig. 6 argue in favor of expansion: for the right panel, since the disk is at a PA near $100°$, the velocities should remain centered on the systemic velocity for any declination in the case of rotation. Likewise, for the upper panel, at the maximum RA offset in east and west, the velocity offset should reach $\pm 2 \, \mathrm{km \, s^{-1}}$ with respect to the systemic velocity, whereas it only reaches $\pm 0.5 \, \mathrm{km \, s^{-1}}$. In brief, in the case of rotation, in Fig. 6, the upper and right panels should be inverted.

Furthermore, for Keplerian rotation, the tangential velocity varies as $1/\sqrt{r}$ and declines toward the stellar systemic velocity, whereas for an expanding disk, it reaches some projected terminal velocity (e.g. Bujarrabal et al. 2005). From the position-velocity diagrams (Figs. 6 & 7) it appears that the velocities do not converge towards the stellar velocity. Finally the CO(1-0) channel maps (Figs. 4 and 5) between 4.2 and $6.2 \, \mathrm{km \, s^{-1}}$ show an extension south, and between 7.0 and $8.6 \, \mathrm{km \, s^{-1}}$, an extension north. This is more easily explained with an expanding disk orthogonal to the north-south outflow. In the following we adopt this hypothesis.

In order to describe the whole structure in more detail, we compare it to a system composed of a biconical flow associated with a flaring disk. The PA of the flaring disk can be further estimated from the RA/velocity and DEC/velocity diagrams in Fig. 6. The amplitudes of the S-shapes are respectively 1 and $4 \, \mathrm{km \, s^{-1}}$ and therefore, the PA is $90 + arctan(1/4) = 104°$, which is consistent with our former estimate ($\sim 100°$) in Sect. 2.2 based on the channel map at $6.6 \, \mathrm{km \, s^{-1}}$ (Fig. 6, lower left). Our simple model with a bipolar flow and an equatorial flaring disk is pictured in Fig. 16. We define the system with 3 angles: $i$, $\epsilon$, and $\phi$, respectively the inclination of the outflow with respect to the plane of the sky, its opening angle, and the opening angle of the flaring disk (as illustrated in Fig. 16).

Since we assume that the bipolar flow is perpendicular to the equatorial disk and that both outflows do not overlap, then $\phi + \epsilon \leq 90°$. A lower limit to the bipolar flow expansion velocity is $8 \, \mathrm{km \, s^{-1}}$ (Table 1), and the maxima of the intensity in the position-velocity diagram (Fig. 6) are obtained for 1 and $12.5 \, \mathrm{km \, s^{-1}}$. The velocity of material along the polar axis projected on the line of sight can be estimated at $5.75 \, \mathrm{km \, s^{-1}}$. The polar axis should thus be inclined with respect to the line-of-sight by at least $arccos(5.75/8.) = 44°$, ie. $i < 46°$.

In the declination-velocity diagramme of Fig. 6, the trace of the disk is clearly S-shaped, with not much emission in the second and fourth quadrants (using the trigonometric convention). It means that, at any declination, the line-of-sight crosses the disk only once (see also Fig. 16), i.e. that $\phi$ cannot be much larger than $i$. However, since in the channel map at $6.6 \, \mathrm{km \, s^{-1}}$, we observe emission from the disk north and south of the star, we need also to assume that $\phi + i \geq 90°$. Therefore, $\phi$ and $i$ should both be ∼ 45°. In the position-velocity diagrams (Fig. 6), the central feature characterizing the disk is well separated from the two high-velocity features, with saddles at ∼ 3.5 km s$^{-1}$ and ∼ 9.5 km s$^{-1}$. Assuming that the bipolar outflow velocity is about 8 km s$^{-1}$, or slightly more, we can estimate the difference between the inclination of the bipolar outflow and its opening angle, $i - \epsilon \geq \arcsin 3/8 = 22°$. Since $i \sim 45°$, we adopt as a compromise $\epsilon \approx 20 - 25°$.

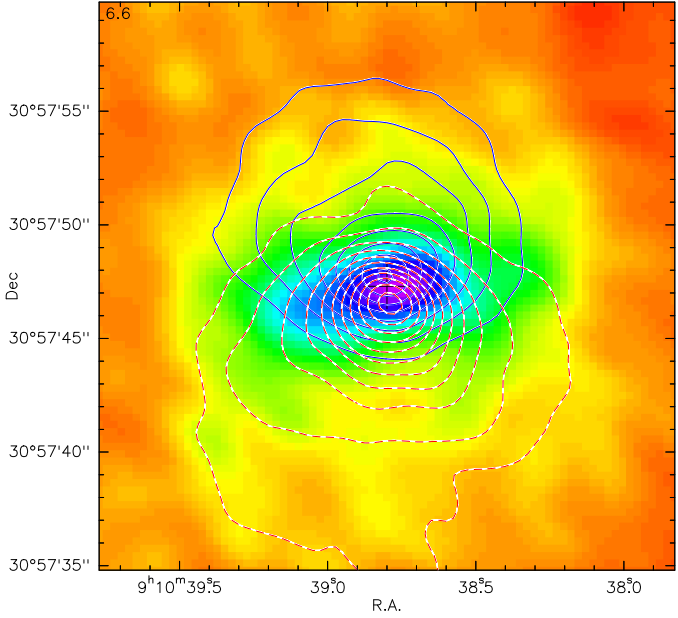

**Fig. 14.** Bipolar structure in CO(1-0). The contour levels range from 0.2 to 2.2 Jy beam$^{-1}$ × km s$^{-1}$ with a step of 0.2 Jy beam$^{-1}$ × km s$^{-1}$. Blue lines: emission integrated over -2 km s$^{-1}$ to 3 km s$^{-1}$. Red dotted lines: emission integrated over 9.5 km s$^{-1}$ to 16 km s$^{-1}$. The background image shows the 6.6 km s$^{-1}$ channel.

In summary, from this qualitative discussion, we conjecture the following parameters for our toy model of CO in RS Cnc: $i \approx 45°$, $\epsilon \approx 20°$, and $\phi \approx 45°$. However, we caution that we have made the hypothesis that all motions are radial. This is probably not correct as micro-turbulent motions of the order of ∼ 0.5 km s$^{-1}$ are commonly assumed in AGB outflows (e.g. Schöier & Olofsson 2001) and as we expect some thermal broadening, possibly of the order of 1 km s$^{-1}$, especially close to the center where CO could be at a temperature ∼ 1000 − 3000 K (e.g. Jura et al. 1988). The inner parts of the disk and jet in Figs. 4 & 5 are not resolved at CO(1-0) and barely at CO(2-1). There may be a hot inner source (of size ≤ 3″, comparable to the beam width in CO(1-0)), extending between 3.5 and 9.5 km s$^{-1}$, possibly related to the emergence of the jet and/or to the interaction of the jet and disk.

Our CO data of high quality will be further analysed with a numerical model that we need to develop (Libert et al., in preparation). This model should take into account the excitation of the CO molecule and the optical depth effects. It will be used to improve our deductions (or will lead us to consider a new geometry). In particular the S-shape morphology that we find in the position-velocity diagram at the declination of the star (Sect. 2.2) should efficiently constrain the modelling of this challenging source.

From the orientation of the structure determined above, we can roughly estimate the mass contained in the bipolar outflow and the disk, based on the CO(1-0) emission. For the disk, we observe a radius of 8″ (see Fig. 6), which, at 122 pc, translates to 1.5 10$^{16}$ cm. With the expansion velocity of ∼ 2.4 km s$^{-1}$ (Table 1), we find a dynamical age for the CO disk of ∼ 2100 yr. With a mass loss rate of 8.6 10$^{-8}$ M$_\odot$ yr$^{-1}$ (Table 1), we estimate a mass of ∼ 1.7 10$^{-4}$ M$_\odot$ yr$^{-1}$.

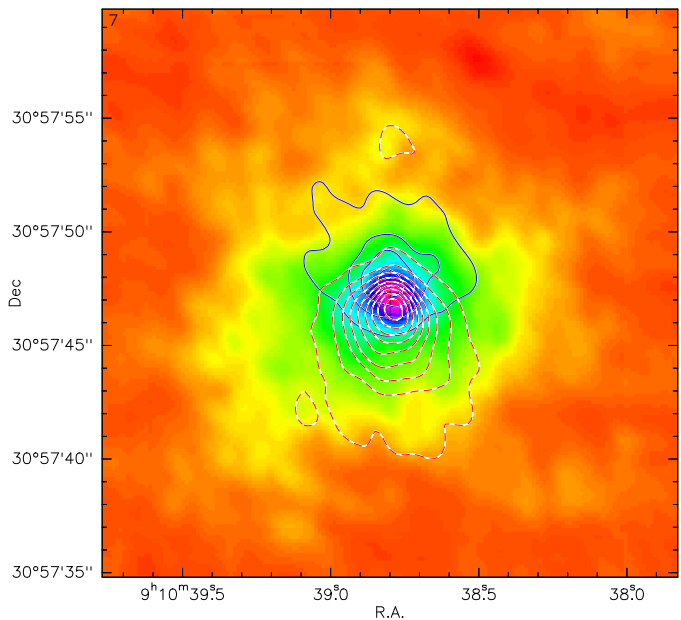

**Fig. 15.** Bipolar structure in CO(2-1). The contour levels range from 0.5 to 6.5 Jy beam$^{-1}$ × km s$^{-1}$ with a step of 0.5 Jy beam$^{-1}$ × km s$^{-1}$. Blue lines: emission integrated over -2 km s$^{-1}$ to 3 km s$^{-1}$. Red dotted lines: emission integrated over 10.5 km s$^{-1}$ to 16 km s$^{-1}$. The background image shows the 7 km s$^{-1}$ channel.

On the DEC/velocity diagrams in Fig. 6, we estimate the extension of the bipolar flow to be about 3″. Given the inclination of 45° deduced above, and the distance of the source, this translates to 7.7 10$^{15}$ cm. For an expansion velocity of 8 km s$^{-1}$ (Table 1), this yields a timescale of ∼ 310 yr. For a mass loss rate between 1 and 4 10$^{-7}$ M$_\odot$ yr$^{-1}$, we estimate the mass contained in the CO bipolar outflow between 0.3 and 1.4 10$^{-4}$ M$_\odot$ yr$^{-1}$.

In both cases, the spatial extent does not exceed the photodissociation radius of the CO molecule (see Table 1). Interestingly, those results are comparable to the estimates of Kahane & Jura (1996), in the case of X Her, for which they claim evidence for a bipolar structure associated with a spherical wind in expansion. However, Nakashima (2005) finds evidence that the kinematical structure traced by the narrow CO component of X Her can be fitted with a Keplerian rotation curve. We do not find such evidence in our data, but we cannot exclude that higher spatial resolution observations might reveal a Keplerian disk in the very center of RS Cnc.

Our H I observations are difficult to interpret, because of the confusion with the ISM lying close in velocity to the intrinsic emission. We find much more extended H I emission than GL2003 and MR2007, leading to an estimate of the hydrogen mass in the RS Cnc envelope of ∼ 3 10$^{-2}$ M$_\odot$, compared to the ∼ 1.5 10$^{-3}$ M$_\odot$ from both GL2003 and MR2007.

However, these results cannot be directly compared, as our estimate takes into account much more distant parts of the RS Cnc envelope, with data obtained up to 44′ to the north (Fig. 13). The H I mass that we derive from an area comparable to the VLA primary beam (30′) is 9.6 10$^{-3}$ M$_\odot$, only ∼ 6 times larger than the estimate of MR2007. Because of the interstellar contamination, there is some uncertainty in the present estimate. However, the discrepancy between the two results is real. It might in part come from a filtering of the source emission by the VLA interferometer. Indeed de Pater et al. (1991) show that, for a source with a Gaussian profile of FWHM = 18′, the actual fraction of the flux detected by the VLA in the D configuration could be only 15 %, in agreement with our factor 6. However, we caution that the emission from RS Cnc is probably not as smooth as a Gaussian profile, which is the worst situation discussed by de Pater et al. In particular, if the envelope has small scale-structure, the VLA would miss much less flux than quoted above.

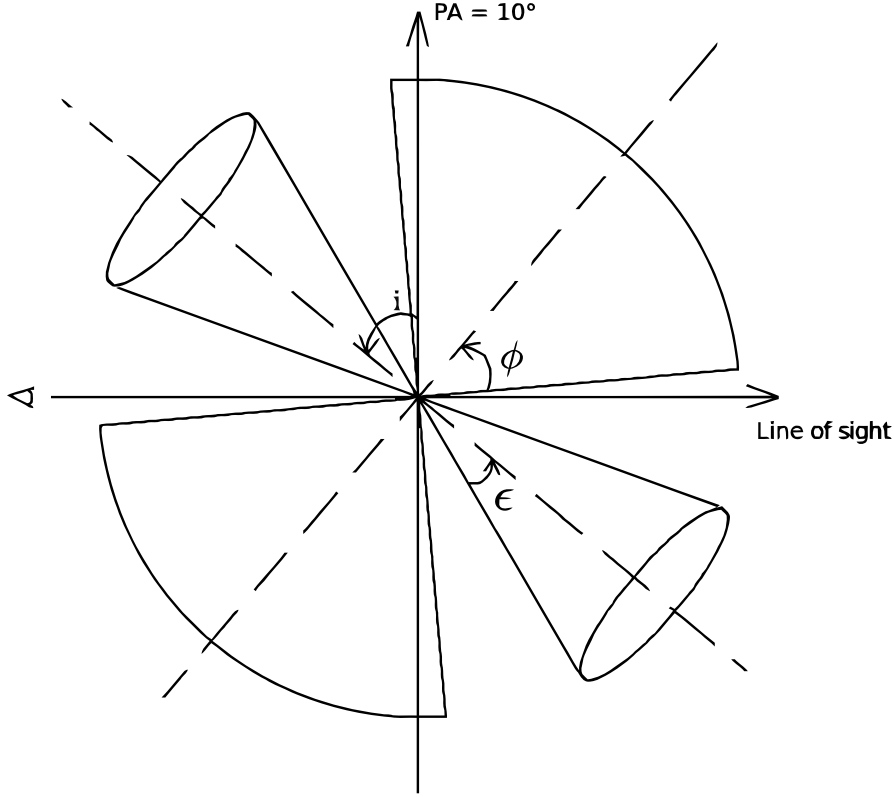

**Fig. 16.** A simple model used to estimate the characteristics of the CO environment of RS Cnc. We represent the plane that contains the line of sight and the polar axis of the structure, inclined at a PA of 10°. $\phi$ is the opening angle of the equatorial disk, $i$ is the inclination of the polar axis with respect to the plane of the sky, and $\epsilon$ is the opening angle of the bipolar flow.

In order to apprehend the genuine contribution of the source, and the extent of its emission, we apply the model already used in GL2006. This model simulates a mass-losing AGB star interacting with an interstellar flow that deforms its shell. It assumes an initially spherical and isotropic wind, at a constant mass-loss rate. To simulate a nonisotropic interaction, the velocity decreases linearly from an inner radius, $r_1$ (expressed in arcminute), following an empirical law $v = v_{exp} - (a + b\cos(\theta))(r - r_1)$, where $\theta$ is the polar angle. This modeled outflow presents a maximum slowing down at $\theta = 0°$ and a minimum at $\theta = 180°$, thus creating an egg-shaped geometry. We orient this "egg" in the plane of the sky with its short dimension ($\theta = 0$) at PA = 135°. In addition, we complete the simulation by accounting for the temperature inside the shell, using the expression adopted for Y CVn, in Paper I: $\log\left(\frac{T}{T_1}\right) = c \log\left(\frac{r}{r_1}\right)$ (where c is a constant and $T_1$ is the temperature at $r_1$). The H I spectral emission of this object can then be evaluated and compared to the observations.

With a constant mass-loss rate of $1\,10^{-7} M_\odot\,yr^{-1}$ (which corresponds to our low estimate from the CO data, cf. Sect. 2.2) and a maximum size of ∼ 30′ at a PA of 315°, we computed the spectra shown in Fig. 17. The parameters are summarized in Table 2. We use a temperature and a density profile that are consistent with our previous studies (Y CVn, Paper I; RX Lep, Paper II). Although the spectra close to the center are well reproduced, we find that the data in the northwest and the southeast positions are not well fitted by the calculated profiles. In the northwest, the problem could be that our simulation does not include the confusion lying under the source (cf. Fig. 12). Indeed, the offset positions at 22′ and 44′ north are too much contaminated by the Galactic H I emission to be reproduced by the simulation. For the southeast problem, the simulation only describes a radial flow whose direction is unaffected by the ISM. Instead, the interaction with the ISM must set in at some distance from the star: the wind material is not only decelerated in the direction of motion, but also swept downstream to form a tail (Wilkin, 1996). This effect will move the center of mass of the H I envelope downstream away from the central star as seen for instance in Mira (Matthews et al., 2008). A more elaborated model, such as the one developed by Villaver et al. (2003), is needed to describe this phenomenon.

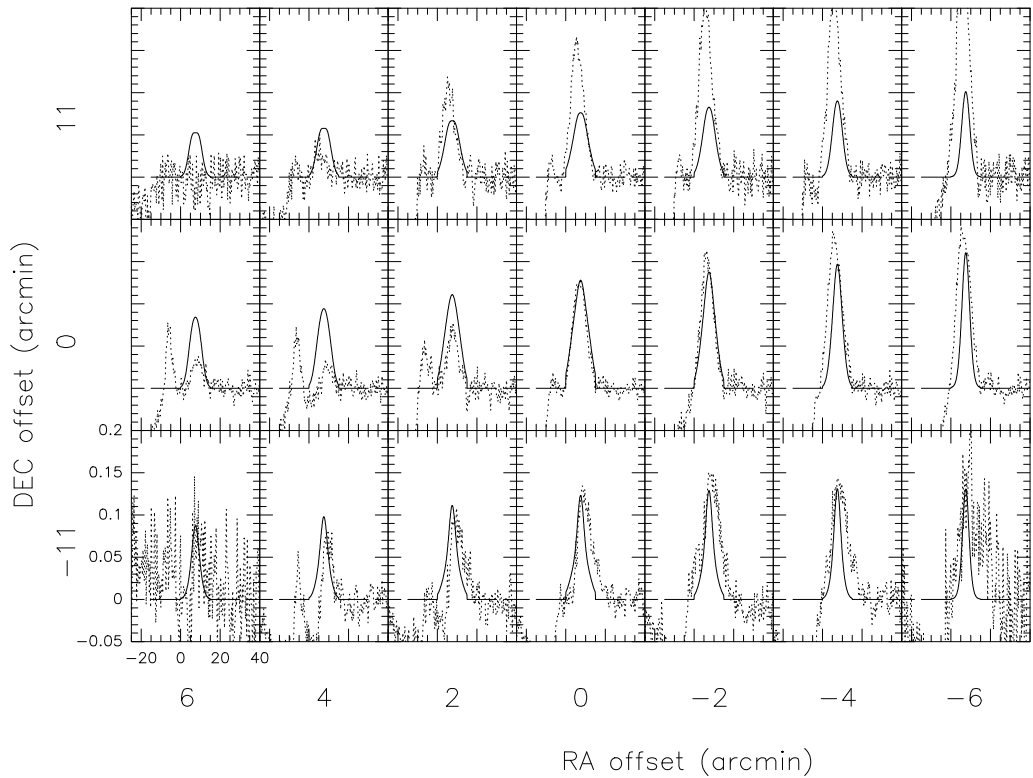

**Fig. 17.** Simulated H I emission of RS Cnc. The solid lines are the results of the model described in Sect. 3, the dotted lines represent the observed profiles.

## 4. Discussion

We find a total circumstellar H I mass of $0.03\,M_\odot$. Nevertheless, it is still a small part of the total mass lost by the star during its evolution. Perhaps some matter lies beyond the limit that we have defined in Fig. 11, and is now indistinguishable from interstellar matter. On the other hand, our mass loss estimate is compatible with that used by Busso & Palmerini (2009) for the present thermal pulse cycle (0.026 M$_\odot$, hence 0.02 M$_\odot$ in H I). However they find a duration for this cycle (5 10$^4$ years) smaller than our age estimate (25 10$^4$ years, Table 2). Therefore our integration might encompass mass lost during several thermal pulse cycles (4 to 5). On the other hand, in that case our present evaluation of the hydrogen mass would then underestimate the mass expelled by RS Cnc during these 4-5 cycles (∼ 0.1 M$_\odot$). We caution that the separation between the genuine H I emission from RS Cnc and the background H I gas is somewhat uncertain and we cannot be sure that the linear interpolation of the Galactic emission behind the source is fully correct. Also, if the stellar effective temperature was lower in the past than now, some hydrogen could still be in molecular form. Observations of the 28 $\mu$m H$_2$ line in the region of the sky that we have delineated in Sect. 2 would be useful.

**Table 2.** Parameters used for the H I modelling and derived quantities.

| | |
|---|---|
| $\dot{M}$ (in atomic hydrogen) | $1 \times 10^{-7}$ M$_\odot$ yr$^{-1}$ |
| M (in atomic hydrogen) | $2.45 \times 10^{-2}$ M$_\odot$ |
| Age | $2.45 \times 10^5$ years |
| PA | 135° |
| $r_1$ | $3.5 \times 10^{-3}$ pc (0.10 ′) |
| $r_2$ ($\theta = 0°$) | 0.46 pc (13 ′) |
| $r_2$ ($\theta = 180°$) | 1.06 pc (30 ′) |
| a | 0.37 s$^{-1}$ |
| b | 0.19 s$^{-1}$ |
| $v_{exp}$ | 7.5 km s$^{-1}$ |
| $v_2$ ($\theta = 0°$) | 0.28 km s$^{-1}$ |
| $v_2$ ($\theta = 180°$) | 2.1 km s$^{-1}$ |
| $T_1$ | 1800 K |
| $T_2$ | 200 K |

Since the bipolar outflow scenario hinted at by the CO observations seems convenient for the case of RS Cnc, it could also be used to revisit the interpretation of EP Aqr, another AGB star showing complex CO winds. Its spectra in CO(1-0) and CO(2-1) present the same type of two component profiles as RS Cnc. Winters et al. (2007) tried to model this emission using a spherical flow with two components: a slow and spatially extended one coupled to a fast and unresolved one. However, a simple two-wind model is not capable of reproducing the shape of the observed spectra along different lines of sight toward the EP Aqr environment. Besides, the model could imply that the fast wind is confined to a region very close to the star (∼ 10$^{15}$ cm), in contradiction with their observations (see their Sect. 5.1). In the case of a bipolar outflow with the polar axis along the line of sight, the two components could probably be reproduced (Libert et al., in preparation). A bipolar outflow may thus be an alternative to explain the composite profiles discovered by K1998. It is worth noting that a similar explanation was already proposed by Kahane & Jura (1996) and also by Nakashima (2005) for X Her, another red giant with composite CO line profiles. Also, Josselin et al. (2000) interpret their spatially resolved CO(2-1) observations of Mira, another source with composite CO line profiles (K1998, Winters et al. 2003), by invoking a spherical envelope disrupted by a bipolar flow.

In summary, there is growing evidence that the composite CO profiles discovered by K1998 are the manifestation of an axisymmetrical structure with a disk and a bipolar outflow, rather than of double winds with different properties.

The direction of the proper motion of RS Cnc, estimated by the Hipparcos measurements, is at a PA of 137° (Sect. 2.1). This direction is opposite to the direction of the H i filament discovered by MR2007 and of the H i structure revealed by the NRT observations, lending support to the idea of an interaction with the surrounding medium around this AGB star. Elongated and deformed circumstellar envelopes have already been observed for several sources in H i. We recently studied the environment of RX Lep (Paper II), which is also an oxygen-rich AGB star, but unlike RS Cnc, it has still not gone through a third dredge-up. For this star, we find an offset H i envelope (by -0.4′ in RA and -4.4′ in DEC) and an elongation (2.3′ of half power width in RA and 15′ in DEC), with a position angle opposed to the direction of the stellar proper motion. We suggested that this deformation might be a consequence of the interaction of the stellar envelope with the interstellar material. Clearly VLA observations of RX Lep would be crucial in order to confirm the H i similarity of the two sources.

An extreme case of such interactions is that of Mira, which presents in H i an envelope assuming a "head-tail" morphology (Matthews et al. 2008). This envelope is consistent on large scale with the far-ultraviolet emission observed by GALEX (Martin et al. 2007) that extends up to 2° north from the star. We investigated the trail with the NRT, observing at several positions along the UV emission, and detected neutral hydrogen at radial velocity decreasing from $45\,\mathrm{km\,s^{-1}}$ (the radial velocity of Mira) as the distance from the star increases. We interpret this result as a stripping of the circumstellar wind by the surrounding medium.

Another evidence for such an interaction with the external medium was discovered in R Hya, by Spitzer. The MIPS image at $70\,\mu$m (Ueta et al. 2006) shows a bow shock structure ahead of the star in the direction of its proper motion. However, no evidence of such structure is found on the IRAS maps of RS Cnc, and Young et al. (1993a) only find an extended emission at $60\,\mu$m that they associate with a shell of 11′ in diameter, centered on the nominal position of the star. The size of the shell is smaller than the one we associate to RS Cnc in H i, but at large distance from the central star, the temperature may not be high enough for the dust emission to be detected or separated from the interstellar one. We also caution that, at this distance, the coupling between gas and dust is unclear and the spatial extension of the dust might differ from that of the gas.

Villaver et al. (2003), and more recently Wareing et al. (2007), have modeled AGB winds interacting with their local environments. Their simulations show that the ram pressure stripping distorts the circumstellar shell as the star moves through the ISM. They concluded that bow shock-like and cometary structures should be common for stars that are moving through the ISM. The increasing number of detections in H i of deformed AGB shells lends support to this prediction.

In general, the morphologies that we find in CO and H i are unrelated. In CO we observe an axisymmetrical structure with a polar orientation at PA $\sim 10°$, whereas in H i we observe a head-tail morphology with a PA at $\sim 310°$. The sizes of the CO and H i regions are very different and this contrast is not unexpected: CO probes the region in which the stellar outflow is emerging and probably shaped by the mass loss process, whereas H i probes the region where the stellar wind is interacting with the ISM and where stellar matter is ultimately injected into it. However, the two structures cannot be completely unrelated and we expect an intermediate region where a transition occurs. In this context it is worth mentionning that MR2007 noted that, in their H i map obtained at the VLA, the compact feature (head) is resolved with a slight elongation along the north-south direction. We have reanalysed their data and show the untapered VLA H i map in Fig. 18 together with the CO(1-0) intensity map. The synthesized VLA beam is about $54''$ (FWHM) and is larger than the CO extension ($\sim 25''$). Based on a 2-D Gaussian fit to the compact feature, we find a deconvolved size of $75'' \times 40''$ at a PA of $14°$. The orientation of the compact H i emission is roughly consistent with that of the CO bipolar outflow. The compact feature is also slightly offset to the north-east, which is consistent with the enhanced CO emission in the same direction. It suggests that in H i we might just be hinting to the transition region. Clearly H i data at higher angular resolution are needed.

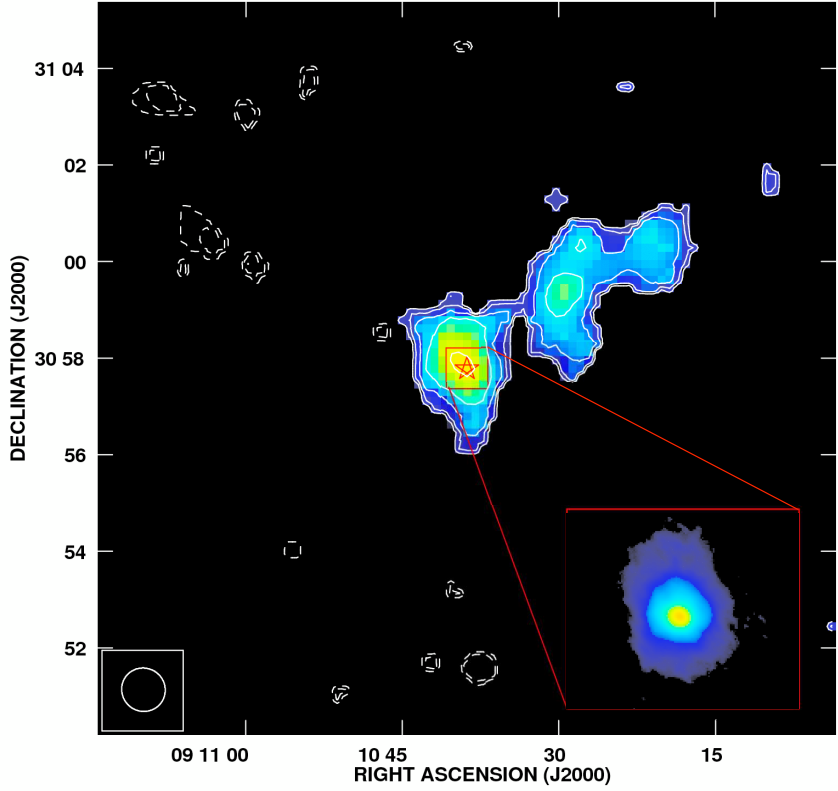

**Fig. 18.** Total intensity map of the H i emission of RS Cnc obtained with naturally weighted, untapered data from the VLA. The contour levels are $0.9 \times (-16,-8,-4,4,8,16,32,64)$ Jy beam$^{-1}$ × m s$^{-1}$ and the intensity levels are 0 to 100 Jy beam$^{-1}$ × m s$^{-1}$. The inset, of size $40'' \times 40''$, is the CO(1-0) emission from PdBI+30 m observations.

## 5. Conclusion

We detect CO and H i emission in the environment of RS Cnc. The CO wind, extending to $12''$, presents a bipolar structure coupled with a velocity gradient oriented along the north-south direction. The H i emission, on the other hand, reveals a completely different structure that is much more extended to the west and the north ($\sim 16'$) than previously observed. This structure has been built up in response to the external environment, and its characteristics do not appear related to those of the internal bipolar structure observed in CO. An average mass loss rate of $\sim 10^{-7}\,M_\odot\,yr^{-1}$, consistent with CO data, over the last 2-3 $10^5$ years can account for this extended structure.

In H I, the NRT reveals a structure which has the same orientation as the filament discovered and mapped with the VLA by MR2007, but which is more extended and more massive. We suspect that one of the causes of this difference is the filtering of extended emission by the VLA, even in the D-configuration (cf. de Pater et al. 1991). It illustrates the power of combining single-dish and interferometric measurements for the study of nearby circumstellar shells, in particular those that are massive and extended.

We emphasize the complementarity of CO and H I observations: with CO, we access the recent events of mass-loss, with emission close to the star, whereas in H I, the more distant parts of the envelope are probed, including the region where it interacts with the ISM. These extended envelopes trace a much older phase of mass loss. They are often deformed and we suspect that these deformations are a frequent phenomenon that might occur for all evolved stars with a significant proper motion with respect to the local medium, as predicted by Villaver et al. (2003).

*Acknowledgements.* The Nançay Radio Observatory is the Unité scientifique de Nançay of the Observatoire de Paris, associated as Unité de Service et de Recherche (USR) No. B704 to the French Centre National de la Recherche Scientifique (CNRS). The Nançay Observatory also gratefully acknowledges the financial support of the Conseil Régional de la Région Centre in France. We are grateful to P. Lespagnol of the NRT staff for handling the 3D data cube. This research has made use of the SIMBAD database, operated at CDS, Strasbourg, France and of the NASA's Astrophysics Data System. We acknowledge useful discussions with M. Busso, R. Guandalini, A. Jorissen, and E. Josselin, and thank the referee and the editor for useful comments. We thank Hans Ungerechts at the 30 m telescope for having obtained the OTF observations, and we are grateful to the IRAM Plateau de Bure staff for their support in the observations.